\documentclass[ secnumarabic, superscriptaddress, amssymb, nobibnotes, aps,prl,twocolumn,longbibliography]{revtex4-1}

\usepackage{amsmath}
\usepackage{graphicx}
\usepackage{mathtools}
\usepackage{xcolor}
\usepackage{todonotes}
\usepackage{hyperref}
\usepackage{gensymb}
\usepackage{ulem}
\usepackage{titlesec}

\titlespacing*{\section}{0pt}{5pt}{5pt}
\titlespacing*{\subsection}{0pt}{9pt}{9pt}

\begin{document}
\titleformat*{\section}{\centering\bfseries\Large}
\titleformat*{\subsection}{\centering\bfseries\large}
\title{Investigating the Electrical Transport Properties and Electronic Structure of Zr\textsubscript{2}CuSb\textsubscript{3}}

\author{Eoghan Downey}
\affiliation{Department of Physics, University of Michigan, Ann Arbor, MI 48109, USA}

\author{Soumya S. Bhat}
\affiliation{Department of Chemical Engineering and Materials Science, Kettering University, Flint, MI, 48504, USA}

\author{Shane Smolenski}
\affiliation{Department of Physics, University of Michigan, Ann Arbor, MI 48109, USA}

\author{Ruiqi Tang}
\affiliation{Department of Physics, University of Michigan, Ann Arbor, MI 48109, USA}

\author{Carly Mistick}
\affiliation{Department of Physics, University of Michigan, Ann Arbor, MI 48109, USA}

\author{Aaron Bostwick}
\affiliation{Advanced Light Source, Lawrence Berkeley National Laboratory, Berkeley, CA 94720, USA}

\author{Chris Jozwiak}
\affiliation{Advanced Light Source, Lawrence Berkeley National Laboratory, Berkeley, CA 94720, USA}

\author{Eli Rotenberg}
\affiliation{Advanced Light Source, Lawrence Berkeley National Laboratory, Berkeley, CA 94720, USA}

\author{Demet Usanmaz}
\altaffiliation{dusanmaz@kettering.edu}
\affiliation{Department of Chemical Engineering and Materials Science, Kettering University, Flint, MI, 48504, USA}

\author{Na Hyun Jo}
\altaffiliation{nhjo@umich.edu}
\affiliation{Department of Physics, University of Michigan, Ann Arbor, MI 48109, USA}

\date{\today}

\def\kill #1{\sout{#1}}
\def\add #1{\textcolor{blue}{#1}} 
\def\addred #1{\textcolor{red}{#1}} 

\begin{abstract}
\noindent
The checkerboard lattice has been proposed to host topological flat bands as a result of destructive interference among its various electronic hopping terms. However, it has proven challenging to realize experimentally due to the difficulty of isolating this structure from any significant out-of-plane bonding while maintaining structural integrity. Here, single crystals of Zr\textsubscript{2}CuSb\textsubscript{3}, a potential candidate for the checkerboard lattice, were synthesized using the solution (self-flux) method, and their structure was confirmed via X-ray diffraction. Electrical transport measurements indicate metallic behavior with electron-dominated carriers. Angle-resolved photoemission spectroscopy reveals multiple electron pockets and significant \textit{k\textsubscript{z}} broadening due to its large \textit{c}-axis and low dispersion features in \textit{k}\textsubscript{z}. Density functional theory calculations further disentangle the contributions from each high-symmetry plane, providing a comprehensive characterization of electronic behavior.
\end{abstract}

\maketitle

\section{I. Introduction}

A ``flat band" refers to a high-density state in which the electron energy dispersion is confined within a narrow range. If the flat band is present near the Fermi level, various intriguing physical phenomena  including unconventional superconductivity and strange metallicity can emerge~\cite{Checkelsky2024}. These phenomena rely on strong correlations among the conduction electrons caused by the reduced bandwidth, which enhances Coulomb interactions over their kinetic energy. 
Flat bands can arise from highly localized orbitals (e.g., 4\textit{f} orbitals), Moir\'e materials, and lattices that promote electronic frustration, such as the Lieb, kagome, or pyrochlore \cite{Neves2024}.
With the addition of spin-orbit coupling, these bands can have non-trivial topological properties and become topological flat bands (TFBs)~\cite{Ma2020}.

\par The recent successful realization of TFBs in the kagome lattice~\cite{Yin2022}, caused by geometric frustration, has prompted the search for the realization of TFBs hosted by other lattice structures \cite{Regnault2022}. One of these lattices, the checkerboard lattice, is defined as the line graph of a square lattice and is made of two square sub-lattices. When the nearest neighbor (NN) and next nearest neighbor (NNN) hopping strengths have the proper ratios, which requires the NNN hopping on one sublatice to be zero, it is expected to potentially host TFBs from the resulting interference~\cite{Katsura2010}\cite{Chang2025}. However, achieving this is challenging in real materials, as crystals often require significant out-of-plane bonding to maintain structural integrity, which destroys the potential for TFBs. Alternatively, a checkerboard lattice can produce TFBs with non-zero NNN hopping in both sub-lattices as long as there are additional next-next-nearest-neighbor (NNNN) hopping terms of the proper strength~\cite{Sun2011}. This creates a more resilient structure and reduces the need for our of plane bonding. However, since accurately calculating the values of these hopping terms \textit{a priori} is very difficult, it is often necessary to synthesize and test candidate materials. In this paper, we investigate crystals of Zr\textsubscript{2}CuSb\textsubscript{3} as a candidate for hosting TFBs. 

\par In Zr\textsubscript{2}CuSb\textsubscript{3} the \textit{ab}-plane has the potential to form one of these checkerboard lattices. Its structure is best understood by comparison to that of Cu\textsubscript{2}Sb \cite{Koblyuk2001}. It is formed by replacing half of one Cu site with Sb and the other Cu site with Zr. Similar compounds of the same structure can also be formed by replacing {Zr} with {Hf} or {Ti} \cite{Koblyuk2003}. This substitution makes the \textit{c}-axis of the crystal longer and helps to isolate the \textit{ab}-planes of its unit cell. If the ratios of the hopping parameters are correct, reducing the bonding between these planes and the rest of the crystal could potentially help checkerboard lattice physics emerge in this system.  

\par While there have been many studies on the electronic structure and properties of Cu\textsubscript{2}Sb, mostly focusing on its extremely large magnetoresistance (MR) values \cite{Endo2021} \cite{Sellmyer1975}, research on Zr\textsubscript{2}CuSb\textsubscript{3} has primarily focused on its crystallography and its polycrystalline transport properties down to 77~K, and it's full electronic structure has not been experimentally measured \cite{Koblyuk2001}\cite{Koblyuk2003}\cite{Melnychenko2003}. 
In this study, we report the synthesis and characterization of single-crystal Zr\textsubscript{2}CuSb\textsubscript{3}, along with its electrical transport properties and electronic structure.
The measured electronic structure was then compared  with the density functional theory ({DFT}) calculations to complete our analysis. 

\section{II. Methods}

\subsection{A. Crystal Growth and Characterization}

\begin{figure}[hbt!]
    \includegraphics [width=\linewidth]{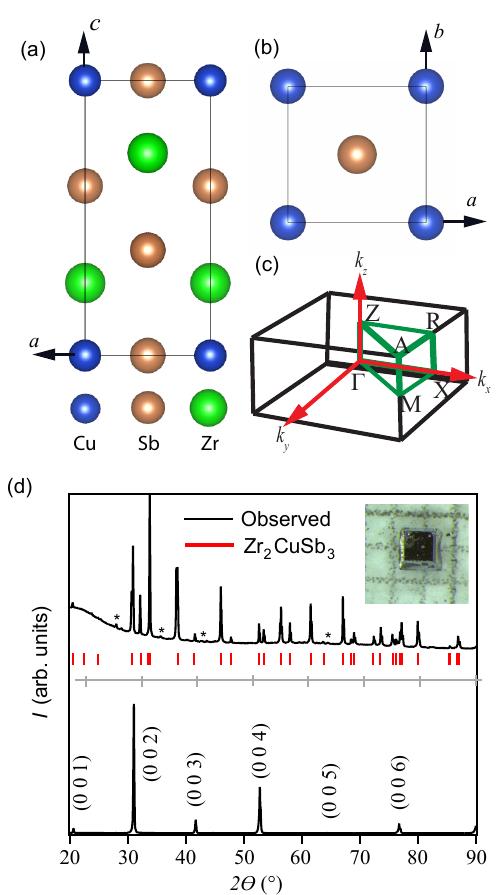}%
  	\caption{(a) Unit cell of Zr\textsubscript{2}CuSb\textsubscript{3} viewed along the \textit{b}-axis with the blue, brown, and green balls representing Cu, Sb, and Zr atoms, respectively. (b) View of the unit cell along the \textit{c}-axis, using the same labels as (a). (c) Illustration of Brillouin zone with high symmetry points and lines marked along with the reciprocal crystallographic directions. (d) {XRD} patterns of the grown crystal, with powder diffraction pattern on top and single crystal diffraction pattern below. The powder XRD pattern is compared to the known Zr\textsubscript{2}CuSb\textsubscript{3}, whose peaks are shown by red lines. A few low intensity peaks, which are marked by asterisks, do not match this pattern and are due to the inclusion of a small amount of Sb flux. The inset shows a picture of a grown single crystal.   
    }
    \label{crystalstr}
\end{figure}

\noindent
High quality single crystals of Zr\textsubscript{2}CuSb\textsubscript{3} were grown using the solution (self flux) method. Elements purchased from Thermo Fisher Scientific were combined in an alumina Canfield crucible with a molar ratio of 5~\% Zr (99.95~\% purity), 37~\% Sb (99.9999~\% purity), and 58~\% Cu (99.9999~\% purity) before being sealed in a quartz tube with Ar under a partial pressure of about 0.33~atm \cite{Canfield2016}\cite{Slade2022}. The crucible was then heated in a furnace to $1000^\circ$C over 5 hours and held at that temperature for another 5 hours. The mixture was then cooled at a constant rate to $740^\circ$C over 90 hours and decanted using a centrifuge~\cite{Canfield1992} at that temperature. The flux growth yielded millimeter-sized single crystals, which form as octahedra with two truncated vertices as shown in the inset of Fig.~\ref{crystalstr}(d). The truncated vertices define the plane perpendicular to the \textit{c}-axis.

\par Samples were ground into powder and analyzed using powder X-ray diffraction (XRD) with Cu $K_{\alpha}$ radiation from a Rigaku Miniflex system~\cite{Jesche2016}. The measured powder pattern shown in Fig.~\ref{crystalstr}(d) matches the known Zr\textsubscript{2}CuSb\textsubscript{3} peaks well. The few peaks that do not match this pattern can be attributed to a small amount of excess Sb flux left in the crystal and have been marked with asterisks in Fig.~\ref{crystalstr}(d). The powder pattern confirms that Zr\textsubscript{2}CuSb\textsubscript{3} crystallizes in a tetragonal structure in the $P-4 m 2$ space group with lattice parameters of $\textit{a}=3.94~\text{\AA}$ and $\textit{c}=8.697~\text{\AA}$ which is in good agreement with previous results on this material~\cite{Koblyuk2001}\cite{Melnychenko2003}. Figures.~\ref{crystalstr}(a) and (b) show a schematic crystal structure of this compound, viewed along the \textit{b} and \textit{c} axes, respectively. Figure.~\ref{crystalstr}(c) shows a schematic of the compound's Brillouin zone, which will guide the electronic structure measurements described later. 
Single crystal {XRD} was also performed using the same equipment and the \textit{c}-axis was clearly identified as shown in the bottom panel of Fig.~\ref{crystalstr}(d). The directions of the electrical transport measurements were based on their relation to this axis and to the plane it defines.
 
\subsection{B. Electrical Transport}

\noindent
Electrical resistance measurements were conducted on polished single crystals of Zr\textsubscript{2}CuSb\textsubscript{3} using a Quantum Design Physical Properties Measuring System (PPMS). Contacts were made using platinum wire with a  diameter of 0.025~mm and H2OE silver epoxy in a standard 4-probe geometry \cite{Miccoli2015}. Measurements were performed using the electrical transport option (ETO) with a 1~mA excitation and an AC frequency of 33.6~Hz. Resistivity was measured with an electrical current applied perpendicular to the \textit{c}-axis from 300~K down to 2~K. Magnetoresistance (MR) was then measured at 2~K in two different geometries. One had current applied perpendicular to the \textit{c}-axis and a magnetic field applied parallel to this axis. The other reversed these directions so that the current was parallel to the \textit{c}-axis and the magnetic field was perpendicular to it. In both geometries the strength of the magnetic field was varied from 0~T to 14~T. Hall measurements were also performed using the same temperature and magnetic field values, but with the contacts arranged in the standard Hall geometry. The magnetic field was applied parallel to the \textit{c}-axis and the current was applied perpendicular to it. 

\subsection{C. Angle-Resolved Photoemission Spectroscopy}

\noindent
 Angle-resolved photoemission spectroscopy (ARPES) measurements were conducted at beamline 7.0.2 (MAESTRO) of the Advanced Light Source at Lawrence Berkeley National Laboratory. The single crystals were mounted on a copper puck and cleaved \textit{in situ} using alumina posts to produce flat, clean surfaces with $k_z$ being aligned parallel to the \textit{c}-axis. Measurements were taken with a beamspot of $\sim$15$\mu $m\,$\times\,15\mu$m at a temperature of $\sim$\,50~K. Horizontally polarized (\textit{p}-polarized) photons with photon energy ranging from $h\nu\,=\,60$\,eV to $h\nu\,=\,150$\,eV were used to probe the $k_z$ dispersion and identify the high symmetry points. Iso-energy and energy dispersion cuts were then taken at $hv=79$~eV ($\Gamma$-X-M plane) and $91$~eV ({Z-A-R} plane). 

\subsection{D. Electronic Structure Calculations}

\noindent
The calculations are performed based on density functional theory (DFT) by using the Vienna Ab-initio Simulation Package ({VASP})~\cite{Kresse_prb_1996, Kresse_compmatsci_1996}.
The Perdew-Burke-Ernzerhof ({PBE}) functional~\cite{Perdew_prl_1996} within the Generalized Gradient Approximation ({GGA}) is employed for the exchange-correlation functional of the electron-electron interactions.
The Projector Augmented Wave ({PAW}) method~\cite{Blchl_prb_1994, Kresse_prb_1999} is applied for the interactions between the core and valence electrons.
A plane-wave kinetic energy cutoff of {414} eV and 13$\times$13$\times$6 $\Gamma$-centered~\cite{Pack_prb_1977} k-points was used for a 6-atom primitive cell.
Geometrical optimizations (cell volume and ionic positions) are performed until the Hellmann–Feynman forces on each atom are smaller than 0.0005 eV/\AA, and the energy convergence criteria are set as 10$^{-8}$ eV.
The valence electron configurations of Cu, Sb, and Zr atoms are considered as 3d$^{10}$ 4p$^1$, 5s$^2$ 5p$^3$, and 4s$^2$ 4p$^6$ 5s$^2$ 4d$^{10}$.
The ground state lattice parameters of $\textit{a}=3.93~\text{\AA}$ and $\textit{c}=8.65~\text{\AA}$ are in good agreement with the experimental results.
Since Zr\textsubscript{2}CuSb\textsubscript{3} contains a heavy {Sb} element, the spin-orbit coupling (SOC) is also included in the electronic structure calculations.
Fermi surface calculations are carried out with the {FermiSurfer} software package~\cite{FermiSurfer}.

\section{III. Results and Discussion}

 \subsection{A. Electrical transport}

\begin{figure}
    	\includegraphics[width=\linewidth]{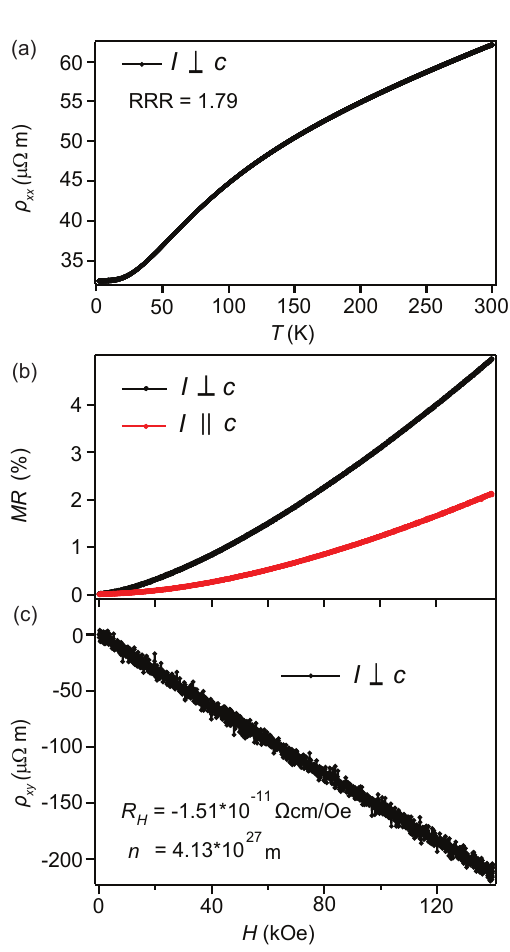}%
  	\caption{(a) Resistivity measured with current applied in the \textit{ab}-plane as a function of temperature. (b) The magnetoresistance is shown as a function of applied magnetic field for both current perpendicular to the \textit{ab}-plane (black curve) and current parallel to it (red curve). The magnetic field was always applied perpendicular to the current. (c) The Hall resistivity was collected as a function of magnetic field applied along the \textit{c}-axis and the current applied perpendicular to it. The Hall coefficient and carrier density are calculated from this data and displayed on the lower left-hand side.
    \label{XRD_XPS_RBS}}
\end{figure}

\begin{figure*}[hbt!]
    	\includegraphics[width=\textwidth]{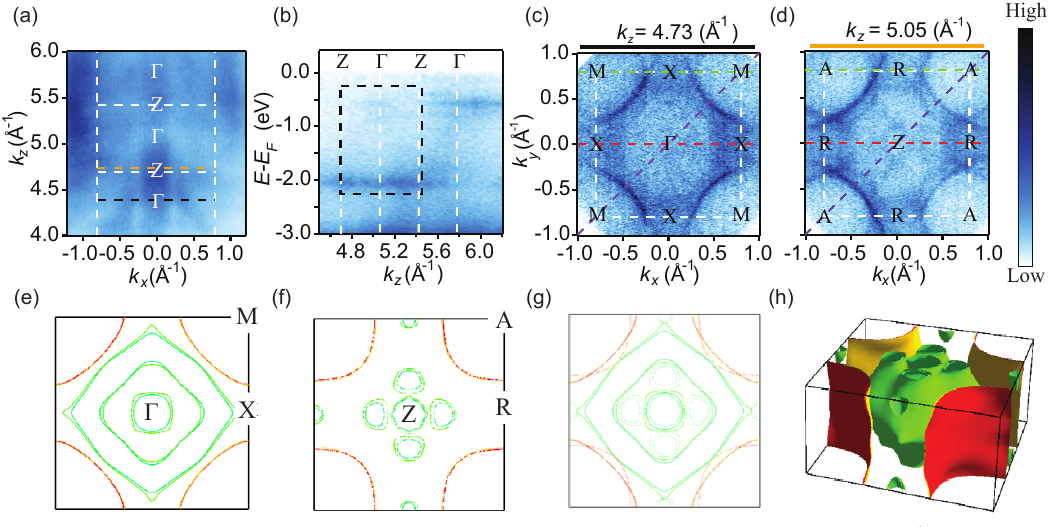}%
  	\caption{Measurements and calculations of the electronic structure of Zr\textsubscript{2}CuSb\textsubscript{3}.(a) The band dispersion in $k_z$ shows two flatter bands at -0.5\,eV and -2~eV, and a dispersive band connecting them. The calculated high symmetry points are marked by white dashed lines, and a black dashed box is used to highlight how the dispersive band lines up with this periodicity. (b) The $k_z$ versus $k_x$ plot. The BZ and high-symmetry points, calculated from the lattice parameters, are indicated by white dashed lines. The energies at which the iso-energy cuts were made are also indicated by black and orange dashed lines. (c) and (d) are ARPES mappings of the Fermi surfaces taken at these marked energies. They correspond to the $\Gamma$-X-M (79eV) and Z-A-R (91eV) planes respectively. (e) and (f) show DFT calculations for the Fermi surfaces of the $\Gamma$ and Z high symmetry planes, while (g) overlays them. (h) shows a calculated 3-dimensional Fermi surface for the whole BZ. 
    \label{Fermi_Surface}}
\end{figure*}

 We begin by measuring the electrical transport properties to test for any strongly correlated effects which could be a sign of a flat band near the Fermi level. Figure~\ref{XRD_XPS_RBS}(a) shows the resistivity of a Zr\textsubscript{2}CuSb\textsubscript{3} crystal as a function of temperature with current applied perpendicular to the \textit{c}-axis. The resistivity shows metallic behavior with no sign of a phase transition down to 2~K and therefore no evidence of correlated effects. The residual resistance ratio ({RRR}) was calculated as the ratio of $\rho$(300~K)/$\rho$(2~K) and found to be 1.79. The results are qualitatively consistent with previous temperature dependent resistivity measurements on polycrystalline Zr\textsubscript{2}CuSb\textsubscript{3}, which were performed down to 77~K \cite{Koblyuk2003}. Small quantitative differences between these two measurements may be due to the averaged anisotropic contributions of the polycrystalline sample and potential differences in sample quality.

The magnetoresitance (MR), defined as $\rho(H)-\rho(H=0))/\rho(H=0)$, is plotted for two different experimental geometries in Fig.~\ref{XRD_XPS_RBS}(b). In the first geometry, shown by the black curve, the current is applied perpendicular to the \textit{c}-axis while the magnetic field is parallel to it. In the second geometry, shown by the red curve, these directions are reversed: the current is applied parallel to the \textit{c}-axis while the magnetic field is applied perpendicular to it. Both configurations exhibit a positive MR that is proportional to $H^2$ and does not saturate at high magnetic fields. However, there is clear anisotropy in the MR. When the current is aligned with the \textit{c}-axis, the MR reaches a maximum of \~2~\% at 14~T. In contrast, when the current is applied perpendicular to the \textit{c}-axis, the MR is more than double, reaching a maximum of \~5~\% at the same magnetic field. This anisotropic behavior indicates differences in the electronic structure, which will be investigated in the following section.

\par We complete our characterization of the electrical transport properties by measuring the Hall effect. Figure~\ref{XRD_XPS_RBS}(c) shows the Hall resistivity measured at 2~K up to a field of 14~T. The magnitude of the Hall resistivity increases linearly with field and shows no change in slope up to 14~T, indicating Zr\textsubscript{2}CuSb\textsubscript{3} is dominated by a single carrier. The slope of this graph gives a Hall coefficient of -1.51$\times$10\textsuperscript{-11}~$\Omega $cm/Oe. We used a simple one band model to calculate the carrier density from this Hall coefficient and found it to be 4.13$\times$10\textsuperscript{27}~m\textsuperscript{-3} electrons. This Hall behavior is best explained by a simple single carrier type model using one parabolic band, suggesting it is unlikely that a TFB is present near the Fermi level.

\subsection{B. Electronic structure} 
\par 

To further characterize the electronic structure of Zr\textsubscript{2}CuSb\textsubscript{3}, ARPES measurements were performed and compared with {DFT} calculations. This will also allow us to look for flat bands that are away from the Fermi level, and therefore would not be detected by the transport probes. Figure~\ref{Fermi_Surface}(a) and (b) show the results of a photon energy scan in terms of \textit{k\textsubscript{z}}. For both plots, we used an inner potential of 1.5~eV, as described in the nearly free final Bloch state model, in order to visually align the features with the known Brillouin zone (BZ) of the crystal \cite{Damascelli2004}. Figure~\ref{Fermi_Surface}(a) shows the electronic dispersion as a function of \textit{k\textsubscript{z}} with a faint dispersive band ranging from -0.5\,eV to -2\,eV. This band, outlined by the dashed box, displays minima and maxima that align with the Z and $\Gamma$ points, respectively, and these features were used to determine the appropriate inner potential.   

\par The dispersive band is fainter and less sharp than the flatter bands at -0.5~eV and -2~eV, due to significant \textit{k\textsubscript{z}} broadening. This broadening arises because momentum is not conserved perpendicular to the surface when electrons escape from the sample, which is known to increase the intensity of bands with no group velocity~\cite{Grandke1978}\cite{Mitsuhashi2016}. We can estimate the amount of \textit{k\textsubscript{z}} broadening with the following equation: $\lambda \cdot \Delta k_{z} \approx 1$, where $\lambda$ is the mean free path of the ejected electrons \cite{Zhang2022}. We took scans using energies of 79-91~eV, giving a probing depth of $ \lambda \approx 5-6~\text{\AA}$  \cite{Seah1979}. This results in $\Delta k_{z} \approx 0.2~\text{\AA}^{-1}$ which is over 50~\% of the distance from the $\Gamma$-{X}-{M}  plane to the {Z}-{R}-{A} plane. This broadening can cause contributions from flatter bands and from high symmetry planes to appear at other measured parts of the BZ, and must be accounted for in our analysis of the electronic structure~\cite{Kumigashira1998}.

 \par The Fermi surfaces taken at the selected photon energies are displayed in Fig.~\ref{Fermi_Surface}(c) and (d), respectively. Both surfaces are outlined with the BZ, the appropriate high symmetry points, and colored dashed lines to indicate where band cuts were taken. These Fermi surfaces are characterized by their similarity, sharing major features. The most obvious is the large electron pocket centered around the {A/M} point, which is actually composed of two different pockets. A small difference appears in the center of the BZ where the {Z} point is surrounded by faint traces of small Fermi pockets, while the $\Gamma$ point is not. Overall these features are well reproduced in the DFT calculations shown in Figs.~\ref{Fermi_Surface}(e) and (f). The Fermi surface calculated for the Z-A-R plane shows the Fermi pockets surrounding the Z point, while the calculations for both planes show larger pockets centered on the A/M points. However, the calculations for these larger pockets do not fully agree with the measured ARPES data. Each plane only shows a singular pocket and they have different shapes, while the measured surfaces show two and are almost identical. This discrepancy can be explained by the \textit{k\textsubscript{z}} broadening, which was mentioned above, causing states from both planes to be measured at both photon energies. This can be most clearly seen in Fig.~\ref{Fermi_Surface}(g) where we have overlapped the two calculations, and the corner pocket now replicates the observed double pocket feature well. The final discrepancy between the observed Fermi surfaces and their calculations is the set of diamond shaped features that are seen in Fig.~\ref{Fermi_Surface}(e) but not in Fig.~\ref{Fermi_Surface}(c). These features may still be present and are possibly obscured by orbital effects or other matrix elements. 

\par
To fully characterize the electronic structure of Zr\textsubscript{2}CuSb\textsubscript{3}, it is desirable to map the Fermi surface over the entire BZ. Since the DFT calculations show excellent agreement with the measured ARPES data at the high symmetry planes, we can be confident in their accuracy throughout the BZ. Therefore, we calculated the Fermi surface of the entire BZ and display the results in Fig.~\ref{Fermi_Surface}(h). The most notable feature is the large cylindrical pockets along the {A}-{M} axis. These pockets show very little dispersion and produce an open Fermi surface along the \textit{k\textsubscript{z}} direction. Open Fermi surfaces are known to produce large magnetoresistance when current is applied along them, as in Cu\textsubscript{2}Sb \cite{Endo2021}\cite{Zhang2019}, but we observe only a small MR effect in this material. The anisotropy of our MR data actually shows a smaller value for current along this direction compared to when it is applied perpendicular to these open pockets. This suggests that, unlike in Cu\textsubscript{2}Sb, Zr\textsubscript{2}CuSb\textsubscript{3} has an additional conduction channel along the \textit{c}-axis which allows the electrons to avoid the high MR of the open Fermi surface. The pockets at the {R} points seem likely candidates for this alternative channel.

\begin{figure*}[hbt!]
    	\includegraphics[width=\linewidth]{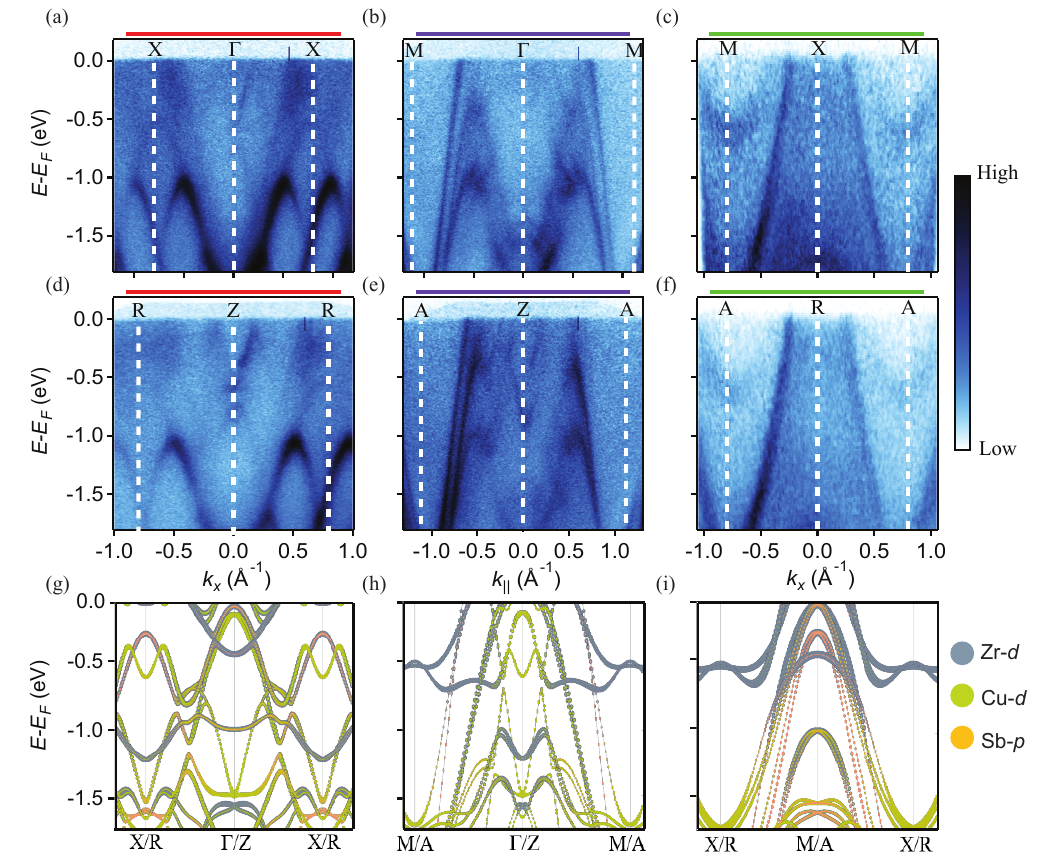}%
  	\caption{Band cuts taken along various high symmetry lines. (a) shows the X-$\Gamma$-X lines, (b) shows the M-$\Gamma$-M line, and (c) shows the M-X-M line. (d), (e) and (f) show band cuts taken along the same directions as (a), (b), and (c), respectively, but in the Z-A-R plane. (g), (h), and (i) show the calculations for both sets of these cuts. Note (a), (b), (d), and (e) all show a small mark around 0.6~\AA\textsuperscript{-1}  near the Fermi level due to a detector error.
    \label{bands}}
\end{figure*}

\par 
Finally, we investigate the band structure of Zr\textsubscript{2}CuSb\textsubscript{3} with band cuts taken along the various high symmetry lines as marked in Fig.~\ref{Fermi_Surface}(c) and (d). As expected, the bands cuts along the same directions in different planes exhibit strong similarities, and the significant \textit{k\textsubscript{z}} broadening causes many bands to appear in both planes. Therefore, our {DFT} calculations for each direction show that both sets of bands overlapped with each other. This overlap gives good agreement with the measured bands and further confirms our understanding of the electronic structure.

\par We examine the agreement between experiment and calculation along the X-$\Gamma$-X/R-Z-R direction shown in Fig.~\ref{bands}(a), (d), and (g). The calculations near the Fermi level reveal pockets surrounding the Z point. Although these features are hard to distinguish in Fig.~\ref{bands}(d), they are present and correspond well to the pockets observed in Fig.~\ref{Fermi_Surface} (d). The experimental band structures also show highly dispersive band between -1.5~eV and -1~eV, which is  well reproduced in the calculations. The band structures along the other two directions in both planes are dominated by large swooping electron bands around the {A/M} point. These bands obviously form the {A/M} pocket discussed previously and are responsible for a large part of the transport behavior of the material. The calculations in Fig.~\ref{bands}(h) and (i) replicate these large bands quite well and show them to come largely from the Zr \textit{d}-orbitals. The M-X-M and A-R-A bands also show a smaller pocket around {A/M}, which reaches down to ~0.5\,eV. These are also represented in the calculations and shown to come from the Zr \textit{d}-orbitals. Overall, the measured band structures are in good agreement with the calculations, and the observed Fermi crossings help confirm the electron dominated nature of Zr\textsubscript{2}CuSb\textsubscript{3}. The band structures also show no flat bands down to -1.5~eV in binding energy.

\section{IV. Conclusion}
\noindent
Single crystals of Zr\textsubscript{2}CuSb\textsubscript{3} were synthesized using the self flux method, and their electronic properties were characterized. We found metallic behavior dominated by a single carrier type in the resistivity with no phase transition down to 2~K. Electronic structure measurements and calculations confirmed this behavior and indicate significant \textit{k\textsubscript{z}} broadening, resulting in electronic states appearing across multiple high symmetry planes. The {DFT} calculations also revealed a cylindrical open Fermi surface along the \textit{k\textsubscript{z}} direction centered at the {A/M} points. Despite this open Fermi surface, no large MR values were observed, and the directional anisotropy showed a reduced MR when current was applied along the direction of these cylindrical surfaces. This indicates that another mechanism is suppressing the MR values. Furthermore, both experimental and theoretical results show no evidence of strongly correlated physics or flat bands in its electronic structure, eliminating Zr\textsubscript{2}CuSb\textsubscript{3} as a candidate for realizing checkerboard lattice physics. 

\section{V. Acknowledgment}
 This material is based upon work supported by the NSF CAREER grant under Award No. DMR-2337535 and under No. DMR-2316831. DU and SB acknowledge support from the NSF LEAPS grant under Grants No. DMR-2316831. This work used resources of the Advanced Light Source, a U.S. Department of Energy (DOE) Office of Science User Facility under Contract No. DE-AC02-05CH11231. 
\pagebreak

\bibliography{reference}

\end{document}